\documentclass[preprint,showpacs,preprintnumbers,amsmath,amssymb]{revtex4}

\usepackage{graphicx}
\usepackage{dcolumn}
\usepackage{bm}
\newcommand{\vecvar}[1]{\mbox{\boldmath$#1$}}
\newcommand{\e}{\mbox{e}}
\newcommand{\erf}{\mbox{erf}}

\begin{document}

\preprint{PRESAT-7901}

\title{Real-space electronic-structure calculations with timesaving double-grid technique}

\author{Tomoya Ono}
\email{ono@upst.eng.osaka-u.ac.jp}
\affiliation{Research Center for Ultra-Precision Science and Technology, Osaka University, Suita, Osaka 565-0871, Japan}
\author{Kikuji Hirose}
\affiliation{Department of Precision Science and Technology, Osaka University, Suita, Osaka 565-0871, Japan}

\date{\today}

\begin{abstract}
We present a set of efficient techniques in first-principles electronic-structure calculations utilizing the real-space finite-difference method. These techniques greatly reduce the overhead for performing integrals that involve norm-conserving pseudopotentials, solving Poisson equations, and treating models which have specific periodicities, while keeping a high degree of accuracy. Since real-space methods are inherently local, they have a lot of advantages in applicability and flexibility compared with the conventional plane-wave approach, and promise to be well suited for large and accurate {\it ab initio} calculations. In order to demonstrate the potential power of these techniques, we present several applications for electronic structure calculations of atoms, molecules and a helical nanotube.
\end{abstract}

\maketitle
\section{INTRODUCTION}
So far, a number of methods for first-principles electronic-structure calculations implemented entirely in real space have been proposed \cite{beck,icp,rs1,rs2,rs3,rs4,rs5,rs6,rs7,rs8}. These have desirable properties compared with the usual plane-wave and/or the linear combination of atomic orbital approaches: (i) Since all of the calculations are carried out in real space, it is easy to incorporate localized Wannier-type orbitals, which are localized in a finite region required for the realization of $O(N)$ calculations, into the algorithm. (ii) A technique utilizing a real-space double-grid \cite{tsdg} is available within the real-space finite-difference (RSFD) formalism, where many grid points are put in the vicinity of nuclei, so that the integrals involving rapidly varying pseudopotentials inside the core regions of atoms can be calculated with a high degree of accuracy. (iii) In order to improve the calculation accuracy, the grid spacing should be narrowed, the procedure of which is simple and definite. Even more important is that (iv) boundary conditions are not constrained to be periodic, e.g., a combination of periodic and nonperiodic boundary conditions for surfaces \cite{dm,jap} and wires \cite{prb-al}, and twist boundary conditions for helical nanotubes are applicable. This aspect (iv) is significantly advantageous to electron-conduction calculations, because an nonperiodic boundary is indispensable for the direction in which electrons flow \cite{icp}.

In this paper, we present a set of efficient techniques for the RSFD approach. In order to demonstrate the applicability of these techniques, we calculate the electronic structures of atoms and molecules by the RSFD approach using isolated boundary condition and that of a $(5,5)$ carbon nanotube utilizing the twist boundary condition. In particular, the real-space calculations with the twist boundary conditions are appropriate for helical nanotubes, since the translational repeat distance leads to extremely large supercells, although the nanotubes can be constructed with one-dimensional translational symmetry. The rest of this paper is organized as follows: in Sec. II, III, and IV, we give in detail the efficient procedures for real-space calculations together with examples, and in Sec. V, we conclude with a discussion on the future direction of the RSFD calculations. The discretization of the Kohn-Sham equation in the RSFD formalism is addressed in Appendix.

\section{Timesaving double-grid technique}
\label{sec:Timesaving double-grid technique}
\subsection{Essential feature of the timesaving double-grid technique}
\label{sec:Essential feature of the timesaving double-grid technique}
In the RSFD formalism, wave functions, electronic charge density, and potentials are all represented on the discrete grid. In general, the discretizations of them are not invariant under uniform translations of the system respect to the position of the grid, which leads a serious problem in practical simulations: the total energy varies unphysically depending on the relative positions of grid points and the nucleus. This problem can be avoided by reducing the grid spacing; however, a fine grid may require so many points as to result in a substantial increase in computational effort. With this as a background, the timesaving double-grid technique was introduced \cite{tsdg} to circumvent this problem easily.

We here extend this technique so that it can be applied to local parts of pseudopotentials as well as nonlocal pseudopotentials. The double-grid consists of two sorts of uniform and equi-interval grid points, i.e., coarse and dense ones, depicted in Fig.~1 by ``$\times$'' and ``$\bullet$'', respectively. The dense-grid region enclosed by the circle is the core region of an atom that is taken to be sufficiently large to contain the cutoff region of pseudopotentials. Throughout this paper we postulate that the wave functions and the charge densities are defined and updated {\it only on coarse-grid points}, while pseudopotentials are strictly given {\it on all dense-grid points} in an analytically or numerically exact manner. In the case of using the pseudopotential parameterized by Bachelet {\it et al.} \cite{hsc} and Troullier and Martins \cite{tmpp} with the Kleinman-Bylander nonlocal form \cite{kb}, the ionic pseudopotential term in the total energy is defined as
\begin{widetext}
\begin{equation}
E = \sum_s \left[ \int_\Omega v^{hard,s}_{loc}(\vecvar{r}-\vecvar{R}^s) \rho(\vecvar{r}) d\vecvar{r} + \int_\Omega v^{soft,s}_{loc}(\vecvar{r}-\vecvar{R}^s) \rho(\vecvar{r}) d\vecvar{r} + \sum_i^M \sum_{lm} \frac{g_{lm,i}^{s*} g_{lm,i}^s}{\langle \psi^{ps,s}_{lm} |\hat{v}_l^s| \psi^{ps,s}_{lm} \rangle} \right],
\end{equation}
\end{widetext}
where $M$ is the number of occupied states and
\begin{equation}
g_{lm,i}^s=\int_\Omega v_{lm}^s(\vecvar{r}-\vecvar{R}^s) \psi_i(\vecvar{r}) d\vecvar{r}.
\end{equation}
Here, $\rho(\vecvar{r})$ is the electron density, and $\vecvar{R}^s$, $v^{hard,s}_{loc}(\vecvar{r})$, and $v^{soft,s}_{loc}(\vecvar{r})$ represent the position of the $s$-th nucleus, the local pseudopotential which varies rapidly around the nucleus (hard local part), and the local pseudopotential which changes gently (soft local part), respectively. The procedure of decomposing the local pseudopotential is explained below. In addition, $v_{lm}^s(\vecvar{r})=\hat{v}_l^s(\vecvar{r}) \psi^{ps,s}_{lm}(\vecvar{r})$, where $\hat{v}_l^s(\vecvar{r})$ and $\psi^{ps,s}_{lm}(\vecvar{r})$ are the nonlocal parts of the pseudopotentials and the pseudowave functions used to prepare the pseudopotentials, respectively. The numerical integrations related to $v^{hard,s}_{loc}(\vecvar{r})$ and $v_{lm}^s(\vecvar{r})$, which vary sharply in the vicinity of nuclei, are performed on the dense grids using the timesaving double-grid technique, whereas the integration related to the gently behaving part $v^{soft,s}_{loc}(\vecvar{r})$ is implemented on the coarse grids. It is noteworthy that $v^{hard,s}_{loc}(\vecvar{r})$ and $v_{lm}^s(\vecvar{r})$ must vanish outside the core region of the pseudopotential; otherwise the calculations of Pulay forces, which are computationally demanding, are required in molecular-dynamics simulations. While the value of the nonlocal part is zero outside the core region, the value of the local part, which is a long-range Coulomb potential, is not zero. Thus, in the case of using the pseudopotential parameterized by Bachelet {\it et al.} \cite{hsc}, $v^{hard,s}_{loc}(\vecvar{r})$ is so set by the following procedure as to be zero outside the core region:
\begin{equation}
v^{hard,s}_{loc}(\vecvar{r}) = -Z^s \left\{\frac{C_1^s \: \erf(\sqrt{\alpha^s_1} |\vecvar{r}|)}{|\vecvar{r}|} - \frac{C_1^s \: \erf(\sqrt{\alpha^s_2} |\vecvar{r}|)}{|\vecvar{r}|} \right\},
\end{equation}
\begin{equation}
v^{soft,s}_{loc}(\vecvar{r}) = -Z^s \left\{\frac{C_1^s \: \erf(\sqrt{\alpha^s_2} |\vecvar{r}|)}{|\vecvar{r}|} + \frac{C_2^s \: \erf(\sqrt{\alpha^s_2} |\vecvar{r}|)}{|\vecvar{r}|} \right\},
\end{equation}
where $\alpha^s_1 \geq \alpha^s_2$ and erf($x$) is the error function (or probability integral).

Let us first consider the inner products between wave functions $\psi$ and nonlocal parts of pseudopotentials $v_{lm}^s$ (see Fig.~2). Here, we assume a one-dimensional case for simplicity. The values of wave functions on coarse grid points ($\circ$) are stored in a computer, and the values on dense-grid points ($\bullet$) are evaluated by interpolating them. The known values of pseudopotentials both on coarse- and dense-grid points ($\circ$) are also shown schematically. Then, from Fig.~2(a) one can see that only the values on coarse-grid points are so inadequate that the inner products can not be accurately calculated; the errors are mainly due to the rapidly varying behavior of pseudopotentials. On the other hand, Fig.~2(b) indicates that the inner products can be evaluated with a high accuracy, if the number of dense-grid points is taken to be sufficiently large and also if the wave functions on dense-grid points are properly interpolated from those on coarse-grid points \cite{comment1}.

There are several interpolation methods for wave functions, among which the simplest is linear interpolation. In this case, the wave functions $\psi_j \equiv \psi(x_j)$ on dense-grid points $x_j$ are interpolated from $\Psi_J \equiv \psi(X_J)$ on coarse-grid points $X_J$ as
\begin{equation}
\label{eqn:linear}
\psi_j = \frac{h - (x_j-X_{J})}h \Psi_{J} + \frac{h - (X_{J+1} -x_j)}h \Psi_{J+1}  ,
\end{equation}
where $h$ is the grid spacing of the coarse-grid points. The inner product is assumed to be accurately approximated by the discrete sum over the dense-grid points, i.e.,
\begin{equation}
\label{eqn:inner}
\int^{d/2+R^s}_{-d/2+R^s} v_{lm}^s(x) \psi(x) dx \approx \sum_{j=-nN_{core}}^{nN_{core}} v_{lm,j}^s \psi_j h_{dens},
\end{equation}
where $v^s_{lm,j} \equiv v_{lm}^s(x_j -R^s)$, $d$ is the ``diameter'' of the core region, $2N_{core}+1$ ($2nN_{core}+1$) is the number of coarse(dense)-grid points in the core region, $h_{dens}$ is the grid spacing of the dense-grid points, and $n=h/h_{dens}$, i.e., $n-1$ is the number of dense-grid points existing between adjacent coarse-grid points. Now, substituting Eq.~(\ref{eqn:linear}) into the right-hand side of Eq.~(\ref{eqn:inner}), we have
\begin{equation}
\label{eqn:kekka}
\int^{d/2+R^s}_{-d/2+R^s} v_{lm}^s(x) \psi(x) dx \approx \sum_{J=-N_{core}}^{N_{core}} w_{lm,J}^s\Psi_J h,
\end{equation}
where
\begin{equation}
w_{lm,J}^s=\sum_{k=-n}^n \frac{h - |x_{nJ+k} - X_J|}{nh}v_{lm,nJ+k}^s.
\end{equation}

Similarly, the inner product of the hard local part and electron density is given by
\begin{eqnarray}
\label{eqn:ono-dg3}
\int^{d/2+R^s}_{-d/2+R^s} v_{loc}^{hard,s}(x) \rho(x) dx &\approx& \sum_{j=-nN_{core}}^{nN_{core}} v^{hard,s}_{loc,j}\rho_j h_{dens} \nonumber \\
&=& \sum_{J=-N_{core}}^{N_{core}} w^s_{loc,J} P_J h, \nonumber \\
\end{eqnarray}
where
\begin{equation}
w^s_{loc,J}=\sum_{k=-n}^n \frac{h - |x_{nJ+k} - X_J|}{nh}v^{hard,s}_{loc,nJ+k} ,
\end{equation}
$v^{hard,s}_{loc,j} \equiv v^{hard,s}_{loc}(x_j -R^s)$, and $P_J$ is the electron density on the coarse-grid point $X_J$.

As shown in Eqs.(\ref{eqn:kekka}) and (\ref{eqn:ono-dg3}), the inner products have been replaced with {\it the summation over coarse-grid points inside the core region}, which produces only a modest overhead in the computational cost. Note that the weight factors $w^s_J$ arising from the interpolation are independent of the wave functions, but dependent only on the known values of pseudopotentials on dense-grid points. Thus, if once computing the factors $w^s_J$ for every molecular-dynamics time step, we do not have to recalculate them throughout self-consistent iteration steps. The extension of the above procedure to the cases of higher-order interpolations is straightforward.

\subsection{Illustration of Double-Grid Efficiency}
\label{sec:Illustration of Double-Grid Efficiency}
We now examine the efficiency of the timesaving double-grid technique by incorporating it into the formalism of the RSFD method. Hereafter, we obey the nine-point finite-difference formula, i.e., $N_f=4$ in Eq.~(\ref{eqn:ono-difference01}), for the differentiation of the wave function. The dense-grid spacing is set as $h^{dens}_\mu=h_\mu/3$, where $h_\mu$ ($\mu=x, y,$ and $z$) is the coarse-grid spacing in the $\mu$ direction. The ninth-order Lagrangian interpolation is used in the dense-grid interpolation. The exchange-correlation effects are treated as the local-spin-density approximation \cite{lda} within the framework of the density-functional theory \cite{hohenberg,kohn}. The pseudopotential database NCPS97 \cite{kobayashi} based on the pseudopotential of Troullier and Martins \cite{tmpp} is used for the ionic potential. The isolated boundary condition is imposed for all directions of $x$, $y$ and $z$.

First, the convergence of total energy for a hydrogen atom as a function of the cutoff energy is depicted in Fig.~3. According to Ref. \cite{gygi}, we defined an equivalent energy cutoff $E_c^{coars}$ [$\equiv (\pi/h_\mu)^2$ Ry] to be equal to that of the plane-wave method which uses a fast-Fourier-transform grid with the same spacing as the present calculation. Hydrogen is one of the most difficult atoms to treat by the RSFD method because the $s$ component of its nonlocal pseudopotential oscillates in the vicinity of the nucleus. The hydrogen atom is placed in the center of the neighboring grid points for the $x$, $y$ and $z$ directions. Without the use of the double-grid technique, the total energy does not converge even when the cutoff energy increases to 17 Ry. On the other hand, when the double-grid technique is employed, the total energy converges even with a small cutoff energy; the total energy converges rapidly and monotonically as the cutoff energy increases.

Next, Fig.~4 shows the total-energy variation as a function of the displacement of the fluorine atom relative to coarse-grid points along a coordinate axis. The coarse-grid spacing is set at 0.33 a.u. The calculation of a model containing a halogen atom requires a quite high cutoff energy because the pseudopotentials of halogen atoms vary rapidly within the core region. Particularly, in the calculation using a discrete grid, such as the RSFD method, the total energy changes in an unphysical manner depending on the relative position of the grid point and the nucleus when the grid spacing is coarse, which prevents us from implementing practical calculations. As seen in Fig.~4, the energy variation in our scheme is $\sim$0.01\% of the total energy (650.4 eV), which is negligibly small.

As the third example, we apply the timesaving double-grid technique to the calculation of the equilibrium bond length of F$_2$ and Cu$_2$ molecules in order to examine the performance of the double-grid method for molecules. The coarse grid spacings are set at 0.33 a.u. for the F$_2$ molecule, and 0.27 a.u. for the Cu$_2$ molecule. Figure~5 shows the adiabatic potential curves for the respective molecules. The results without any interpolation have many ``humps'', and the distances between adjacent humps on the respective dotted curves are close to twice the length of the grid spacing $h_\mu$, which confirms that the oscillation depends on the relative position of the atom with respect to coarse-grid points. To circumvent this problem, Gygi {\it et al.} \cite{gygi} proposed a real-space grid in adaptive curvilinear coordinates. However, owing to the use of these distorted coordinates, their scheme needs {\it Pulay forces} in molecular-dynamics simulations; this is computationally very demanding. On the contrary, our approach requires only the Hellmann-Feynman forces, which significantly reduces the computational cost concerning the calculation of forces. As shown in Fig.~5, the locations of the minima in the result using the double-grid technique agree with the equilibrium bond length (2.68 a.u. for F$_2$ and 4.19 a.u. for Cu$_2$) obtained by experiments and other methods. These results clarify that the present method solves this problem and is very efficient and applicable for practical simulations.

\section{Fast solver for Poisson equation}
\subsection{Fuzzy cell decomposition and multipole expansion technique}
\label{sec:fuzzy cell decomposition and multipole expansion technique}
The Hartree potential in the Kohn-Sham equation is commonly evaluated by solving the Poisson equation
\begin{equation}
\nabla^2v_H(\vecvar{r})=-4\pi\rho(\vecvar{r}).
\end{equation}
In the case of isolated boundary conditions, one needs to determine the boundary values of the Hartree potential just outside of the calculation domain to set up the matrix for solving the Poisson equation \cite{comment2}. For small systems, the numerical summation over the grid is the most direct procedure. On the other hand, when the system becomes large, the numerical summation requires so many operations as to result in a demanding scheme, since the computational cost for the direct summation is proportioned to the 5/3 power of the model size. Chelikowsky {\it et al.} \cite{chelikowsky1, chelikowsky2} proposed the procedure using a multipole expansion of the charge density around an arbitrary point $\vecvar{r}^0$.
\begin{eqnarray}
v_H(\vecvar{r})&=& \int \frac{\rho(\vecvar{r}')}{|\vecvar{r}-\vecvar{r}'|} d\vecvar{r}' \nonumber \\
&=&\sum_{l=0}^\infty \int \frac{\rho(\vecvar{r}')}{|\vecvar{r}-\vecvar{r}^0|} \left(\frac{|\vecvar{r}'-\vecvar{r}^0|}{|\vecvar{r}-\vecvar{r}^0|}\right)^lP_l(\cos \theta') d \vecvar{r}' \nonumber \\
&=&\frac{\int \rho(\vecvar{r}') d \vecvar{r}'}{|\vecvar{r}-\vecvar{r}^0|}+\sum_{\mu} p_\mu \cdot \frac{(r_\mu-r_\mu^0)}{|\vecvar{r}-\vecvar{r}^0|^3} \nonumber \\
&& + \sum_{\mu,\nu} q_{\mu \nu} \cdot \frac{3\:(r_\mu-r_\mu^0)(r_\nu-r_\nu^0)-\delta_{\mu \nu}|\vecvar{r}-\vecvar{r}^0|^2}{|\vecvar{r}-\vecvar{r}^0|^5} \nonumber \\
&& + \cdots
\label{eqn:poisson01}
\end{eqnarray}
Here, $\mu$ and $\nu$ denote $x$, $y$ and $z$. In addition, the functions $P_l(\cos \theta')$ $(l=0, 1, 2, \cdots)$ are the Legendre polynomials and $\cos \theta '$ is expressed as
\begin{equation}
\cos \theta'=\frac{(\vecvar{r}-\vecvar{r}^0) \cdot (\vecvar{r}'-\vecvar{r}^0)}{|\vecvar{r}-\vecvar{r}^0| \cdot |\vecvar{r}'-\vecvar{r}^0|},
\end{equation}
$p_\mu$ is the dipole moment,
\begin{equation}
p_\mu=\int (r_\mu'-r_\mu^0) \rho(\vecvar{r}') d \vecvar{r}',
\end{equation}
and $q_{\mu \nu}$ is similar to quadrupole moment \cite{comment},
\begin{equation}
q_{\mu \nu}=\int \frac{1}{2} (r_\mu'-r_\mu^0)(r_\nu'-r_\nu^0) \rho(\vecvar{r}') d \vecvar{r}' .
\end{equation}
By virtue of this method, one can make the computational cost proportioned to the model size. However, the accuracy of the solution largely depends on the choice of the position $\vecvar{r}^0$ because the expansion is carried out only around one point.

We now introduce the fuzzy cell decomposition and multipole expansion (FCD-MPE) method, which is free from this problem. A weighting function $\omega_s(\vecvar{r})$ for the multiple-center system centered at the $s$-th nucleus is prepared. This function is the defining function of so-called Voronoi polyhedra $\Omega_s$ \cite{voronoi}, which provides Wigner-Seitz cells and is set to satisfy the following equations.
\begin{equation}
\label{eqn:ono-poisson08}
\sum_s \omega_s(\vecvar{r}) = 1
\end{equation}
\begin{equation}
\label{eqn:ono-poisson09}
\omega_s(\vecvar{r}) = \left \{
\begin{array}{ll}
1 & \hspace {5mm} \in \Omega_s \\
0 & \hspace {5mm} \mbox{otherwise}
\end{array}
\right.
\end{equation}
The charge density is divided into the charge distribution existing around the $s$-th nucleus using the weighting function $\omega_s(\vecvar{r})$.
\begin{equation}
\rho_s(\vecvar{r})=\rho(\vecvar{r})\omega_s(\vecvar{r}),
\end{equation}
\begin{equation}
\rho(\vecvar{r})=\sum_s\rho_s(\vecvar{r})
\end{equation}
By performing the multipole expansion for each $\rho_s(\vecvar{r})$ centered around the position of each nucleus $\vecvar{R}^s$, we have
\begin{eqnarray}
\label{eqn:ono-poisson10}
v_H(\vecvar{r}) &=& \sum_s \Biggl( \frac{\int \rho_s(\vecvar{r}') d \vecvar{r}'}{|\vecvar{r}-\vecvar{R}^s|} + \sum_{\mu} p_\mu^s \cdot \frac{r_\mu-R^s_\mu}{|\vecvar{r}-\vecvar{R}^s|^3} \nonumber \\
&&+ \sum_{\mu,\nu} q_{\mu \nu}^s \cdot \frac{3(r_\mu-R^s_\mu)(r_\nu-R^s_\nu)- \delta_{\mu \nu}|\vecvar{r}-\vecvar{R}^s|^2}{|\vecvar{r}-\vecvar{R}^s|^5} \nonumber \\
&&+ \cdots \Biggr) , \nonumber \\
\end{eqnarray}
where $\mu$ and $\nu$ are $x$, $y$ and $z$. In addition, $p_\mu^s$ and $q_{\mu \nu}^s$ are
\begin{equation}
p_\mu^s=\int (r_\mu'-R_\mu^s) \rho_s(\vecvar{r}') d \vecvar{r}',
\end{equation}
and
\begin{equation}
q_{\mu \nu}^s=\int \frac{1}{2} (r_\mu'-R_\mu^s)(r_\nu'-R_\nu^s) \rho_s(\vecvar{r}') d \vecvar{r}' .
\end{equation}

Note that if $\omega_s(\vecvar{r})$ were in the manner of a step function at the boundary of $\Omega_s$, many terms would be required for the expansion of Eq.~(\ref{eqn:ono-poisson10}). In order to carry out the expansion with as small a number of terms as possible, the behavior of $\omega_s(\vecvar{r})$ near the boundary should be made as smooth as possible using the fuzzy cell technique \cite{fuzzy}, in which the section of the boundary of Eq.~(\ref{eqn:ono-poisson09}) is fuzzy. When the fuzzy cell is employed, the multipole expansion up to the quadrupole is sufficient to obtain an accurate solution.

\subsection{Efficiency of fuzzy cell decomposition and multipole expansion method}
\label{sec:Efficiency of the fuzzy cell decomposition and multipole expansion method}
As a numerical test of the the FCD-MPE method in accuracy, the Poisson equation is solved employing the boundary values obtained by the FCD-MPE method and those by the multipole expansion around one point in the calculation domain. Here, the charge distribution is assumed to be
\begin{equation}
\rho(\vecvar{r})= \sum_{i=1,2} \gamma_{1,i} \times \left( \frac{\gamma_{2,i}}{\pi} \right)^{\frac{3}{2}} \exp[{-\gamma_{2,i}(\vecvar{r}-\vecvar{a}_i)^2}], 
\end{equation}
which imitates the charge distribution of a diatomic molecule, where $\vecvar{a}_1=(2,0,0)$, $\vecvar{a}_2=(-2,0,0)$, $\gamma_{1,1}=6$, $\gamma_{1,2}=4$, $\gamma_{2,1}=0.8$, and $\gamma_{2,2}=0.6$. The cell size is set at $18.0\times 16.0\times 16.0$ a.u., a nine-point finite difference formula, i.e., $N_f=4$ in Eq.~(\ref{eqn:ono-difference01}), is adopted for the second-order derivative, and the grid spacing is chosen to be 0.15 a.u. which is smaller than those of practical calculations so as to suppress the numerical error due to the finite-difference approximation. In the case of using Eq.~(\ref{eqn:poisson01}), the expansion is implemented around
\begin{equation}
\vecvar{r}^0=\frac{\int \vecvar{r}'\rho(\vecvar{r}') d \vecvar{r}'}{\int \rho(\vecvar{r}') d \vecvar{r}'}.
\end{equation}

Figure~6 shows the relative errors between the values of the Hartree potential for the cross section at $z=0.075$ a.u. using the computed boundary values and the analytical boundary value, i.e.,
\begin{equation}
v_H(\vecvar{r})= \sum_{i=1,2} \gamma_{1,i} \times \frac{\erf \,(\sqrt{\gamma_{2,i}}(\vecvar{r}-\vecvar{a}_i))}{|\vecvar{r}-\vecvar{a}_i|}.
\end{equation}
As is evident from these results, the accuracy of the calculation is improved markedly by the FCD-MPE method.

\section{Kohn-Sham Hamiltonian under twist boundary condition}
\label{sec:ono-Kohn-Sham Hamiltonian under twist boundary condition}
\subsection{Discretization of screw operator for twist boundary condition}
We here address the discretization of the Kohn-Sham equation under the twist boundary condition (see Fig.~7 for an example). The boundary-condition operators in $x$ and $y$ directions, $\tau_x$ and $T_y$ (refer to Appendix), are zero and the operator in $z$ direction, $T_z$, is represented by using a screw operator $\mathcal{S}$ \cite{mintmire1}, which is an $N_{xy}(=N_x \times N_y)$-dimensional matrix in terms of a translation $L_z$ down (up) the $z$ axis in conjunction with a right-handed rotation $\varphi$ ($-\varphi$) about the $z$ axis. Here, $N_x$ and $N_y$ are the numbers of grid points in the $x$ and $y$ directions, respectively.

Let us consider the discretization of the screw operator $\mathcal{S}$. The wave function multiplied by the screw operator is defined as
\begin{eqnarray}
| \Psi'(z_k) \rangle &\equiv& \mathcal{S}(\varphi) | \Psi(z_k) \rangle \nonumber \\
&\approx& \mathcal{S}_d(\varphi) | \Psi(z_k) \rangle,
\end{eqnarray}
where $\Psi(z_k)$ is a columnar vector consisting of $N_{xy}$ values of the wave function on the $x$-$y$ plane at the $z=z_k$ point, $\psi_{z_k}(x_i,y_j)$. Note that the wave function $\psi'_{z_k}$ cannot be evaluated exactly in the real-space formalism with the Euclid coordinate, since the twist angle $\varphi$ is not always in a multiple number of $\pi/2$. Therefore, the wave function is obtained by an interpolation using the values of the wave function on the neighboring grid points. In the case of the linear interpolation, we have
\begin{eqnarray}
\psi'_{z_k}(x_i,y_j) & = & \xi_1 \psi_{z_k}(x_{i'}, y_{j'}) + \xi_2 \psi_{z_k}(x_{i'+1}, y_{j'}) \nonumber \\
&&+ \xi_3 \psi_{z_k}(x_{i'}, y_{j'+1}) + \xi_4 \psi_{z_k}(x_{i'+1}, y_{j'+1}), \nonumber \\
\end{eqnarray}
where
\begin{eqnarray}
\xi_1 &=& (x_{i'+1} - x_i \cos \varphi + y_j \sin \varphi)/h_x \nonumber \\
&& \times (y_{j'+1} - x_i \sin \varphi - y_j \cos \varphi)/h_y, \nonumber \\
\xi_2 &=& (x_i \cos \varphi - y_j \sin \varphi - x_{i'})/h_x \nonumber \\
&& \times (y_{j'+1} - x_i \sin \varphi - y_j \cos \varphi)/h_y, \nonumber \\
\xi_3 &=& (x_{i'+1} - x_i \cos \varphi + y_j \sin \varphi)/h_x \nonumber \\
&& \times (x_i \sin \varphi + y_j \cos \varphi - y_{j'})/h_y, \nonumber \\
\xi_4 &=& (x_i \cos \varphi - y_j \sin \varphi - x_{i'})/h_x \nonumber \\
&& \times (x_i \sin \varphi + y_j \cos \varphi - y_{j'})/h_y.
\end{eqnarray}
Here, $x_{i'}$, $x_{i'+1}$, $y_{j'}$ and $y_{j'+1}$ are so determined as to satisfy $x_{i'} \leq x_i \cos \varphi - y_j \sin \varphi < x_{i'+1}$ and $y_{j'} \leq x_i \sin \varphi + y_j \cos \varphi < y_{j'+1}$. Whereas the exact screw operator holds $\mathcal{S}(-\varphi)=\mathcal{S}^\dagger(\varphi)$, the discretized screw operator does not necessarily satisfy the identity due to the loss of accuracy of the interpolation using discrete grid points. The boundary condition $T_z$ in Eq.~(\ref{difference01}) is defined as $\e^{ik_zLz}[\mathcal{S}_d(\varphi) + \mathcal{S}_d^\dagger(-\varphi)]/2$ so as to keep the discretized Hamiltonian as an Hermitian, where $k_z$ is the Bloch wave number. Concerning the long-range ionic potential in the case of the twist boundary condition, refer to Ref. \cite{icp}.

\subsection{Application: Band structure of helical carbon nanotube}
Carbon nanotubes \cite{iijima} consist of a few concentric tubes each of which has carbon-atom hexagons arranged in a helical fashion about the axis. Their topologies are characterized by a set of two numbers $(n, m)$ called chiral indices. Since they were first discovered, their atomic and electronic structures have been intensively explored by both experimental and theoretical approaches. Previous theoretical studies have shown that the carbon nanotubes change between having metal and semiconductor properties depending on their structure such as their diameter and helical arrangement \cite{hamada,saito}. In order to present the efficiency and applicability of the RSFD method, we compute the electronic band structure of the $(5,5)$ carbon nanotube using the twist boundary condition.

The computational conditions are as follows: the influences of core electrons are induced by the Troullier-Martins-type pseudopotential \cite{kobayashi,tmpp} in the Kleinman-Bylander nonlocal form \cite{kb}. The tube structure is generated with a radius of 6.41 a.u., a screw operation having a twist angle of $\pi /5$ rad and a translational shift of $L_z$=2.32 a.u. chosen to yield nearest-neighbor separations between rings equal to the in-ring values. Exchange-correlation effects are treated using the local density approximation \cite{lda} and the central finite-difference formula, i.e., $N_f=1$ in Eq.~(\ref{eqn:ono-difference01}), is adopted. The coarse grid spacing is set at 0.33 a.u. and the denser grid spacing is set at 0.11 a.u. in the vicinity of the nuclei with the augmentation of double-grid points.

We show in Fig.~8(a) the helical band structure of the $(5,5)$ carbon nanotube. Since there are multiple equivalent ways of depicting the band structure of ($n, n$) carbon nanotubes, Mintmire {\it et al.} \cite{mintmire2} proposed a pseudoband structure. By their procedure, we can classify all the bands into two representations; an ``$a$'' representation which varies in the Brillouin zone depending on possible helical operators and an ``$e$'' representation which moves in the Brillouin zone depending on helical twist angle. The pseudoband structure is obtained by shifting all the ``$e$'' bands in $k$ by a quantity corresponding to a phase factor of $\pm j\varphi$. This choice of phase shifts the ``$e$'' bands to doubly degenerate bands.

The pseudoband structure of the $(5, 5)$ carbon naotube is illustrated in Fig.~8(b). We also depict in Fig.~8(c) the band structure using the conventional supercell which has twice the length in the tube direction. One can recognize that the present pseudoband structure leads to a depiction of the bands as equivalent as possible to the bands that would be present if the system had translational periodicity with the repeat distance of the helical supercell.

\section{SUMMARY}
We have presented efficient techniques for the RSFD calculations together with several examples. The main points of our method are as follows: (i) The timesaving double-grid method can suppress the spurious oscillation of the total energy for the displacement of the atom relative to the coarse-grid points, and there is little computational effort for the double-grid method by virtue of the integration over the coarse-grid points. (ii) The FCD-MPE method improves the accuracy of the computation for the boundary values of the Hartree potential required in the case of the isolated boundary condition. (iii) We can easily treat a system having a twist angle at its boundary by including a twist operator in the Kohn-Sham Hamiltonian. The pseudoband structure of the $(5,5)$ carbon nanotube agrees with that calculated employing the large supercell.

From what has been discussed above, our techniques enable us to implement highly accurate first-principles calculations based on the RSFD approach and are efficiently applicable to various systems, in particular, a system having a twist angle at its boundary. When coupled with conductance calculations, this approach would be useful in studying electron transport behavior in helical nanotubes \cite{icp,hirose2}.

\section*{ACKNOWLEDGMENTS}
This research was supported by a Grant-in-Aid for the 21st Century COE ``Center for Atomistic Fabrication Technology'' and also by a Grant-in-Aid for Scientific Research (C) (Grant No. 16605006) from the Ministry of Education, Culture, Sports, Science and Technology. The numerical calculation was carried out by the computer facilities at the Institute for Solid State Physics at the University of Tokyo, and the Information Synergy Center at Tohoku University.

\appendix
\section{Kohn-Sham equation in real-space finite-difference formalism}
We demonstrate the procedure for discretizing the Kohn-Sham Hamiltonian \cite{hohenberg,kohn} in the real-space formalism. In the conventional methods using basis sets, the derivative of wave functions arising from the kinetic-energy operator in the Kohn-Sham Hamiltonian can be computed by differentiating the basis. However, since the prime concept of the RSFD method does not employ any bases, the second derivative of function $f$($x$) at grid point $x=ih_x$ ($i$: integer) is approximated by the following finite-difference formula.
\begin{equation}
\label{eqn:ono-difference01}
\left. \frac{d^2}{dx^2} f(x) \right|_{x=ih_x} \approx \sum_{n=-N_f}^{N_f} c_n f(ih_x+nh_x)
\end{equation}
Here, $N_f$ and $h_x$ are the parameters determining the order of the finite-difference approximation and the grid spacing, respectively. In the case of $N_f=1$, the coefficients are $c_{-1}=c_1=1/h_x^2$, and $c_0=-2/h_x^2$. See Ref. \cite{chelikowsky2} for the values of coefficients $c_n$ when $N_f>1$. Although we here present the case of the central finite difference, i.e., $N_f=1$, and local pseudopotentials in order to describe the essence of this scheme, the inclusions of higher-order finite-difference formulas and nonlocal parts of the norm-conserving pseudopotentials are straightforward.

In the RSFD method, the wave functions and electron density are given only on discrete grid points in three-dimensional real space. Therefore, the Kohn-Sham Hamiltonian \cite{kohn} acting on a wave function must be given in a form discretized in real space. The discretized Kohn-Sham equation is written as
\begin{widetext}
\begin{equation}
\label{difference01}
\left[
\begin{array}{ccccccc}
V(z_1) & B_z & 0 & 0 & \cdots & 0 & T_z^\dagger \\
B_z^\dagger & V(z_2) & B_z & 0 & & & 0 \\
0 & \ddots & \ddots & \ddots & \ddots & & \vdots \\
\vdots & \ddots & B_z^\dagger & V(z_k) & B_z & \ddots & \vdots \\
\vdots & & \ddots & \ddots & \ddots & \ddots & 0 \\
0 & &  & 0 & B_z^\dagger & V(z_{N_z-1}) & B_z \\
T_z & 0 & \cdots & 0 & 0 & B_z^\dagger & V(z_{N_z})
\end{array}
\right]
\left[
\begin{array}{c}
\Psi(z_1) \\ \Psi(z_2) \\ \vdots \\ \Psi(z_k) \\ \vdots \\ \Psi(z_{N_z-1}) \\ \Psi(z_{N_z})
\end{array}
\right] = E \left[
\begin{array}{c}
\Psi(z_1) \\ \Psi(z_2) \\ \vdots \\ \Psi(z_k) \\ \vdots \\ \Psi(z_{N_z-1}) \\ \Psi(z_{N_z})
\end{array}
\right],
\end{equation}
\end{widetext}
where $T_z$ is an $N_{xy}(=N_x \times N_y)$-dimensional matrix specified by the boundary condition in the $z$ direction [for example, see Eq.(\ref{eqn:zbound})], $\Psi(z_k)$ is a columnar vector consisting of $N_{xy}$ values of the wave function on the $x$-$y$ plane at the $z=z_k$ point, $\psi_{z_k}(x_i,y_j)$, $B_z$ is a constant matrix proportional to $N_{xy}$-dimensional unit matrix $I$,
\begin{equation}
B_z=-\frac{1}{2h_z^2}I,
\end{equation}
and $N_x$, $N_y$ and $N_z$ are the numbers of grid points in the $x$, $y$ and $z$ directions, respectively. In addition, $V(z_k)$ is the following $N_{xy}$-dimensional matrix defined on the $x$-$y$ plane at the $z=z_k$ point.
\begin{widetext}
\begin{equation}
\label{difference02}
V(z_k) = \left[
\begin{array}{ccccccc}
V_{z_k}(y_1) & B_y & 0 & 0 & \cdots & 0 & T_y^\dagger \\
B_y^\dagger & V_{z_k}(y_2) & B_y & 0 & & & 0 \\
0 & \ddots & \ddots & \ddots & \ddots & & \vdots \\
\vdots & \ddots & B_y^\dagger & V_{z_k}(y_j) & B_y & \ddots & \vdots \\
\vdots & & \ddots & \ddots & \ddots & \ddots & 0 \\
0 & & & 0 & B_y^\dagger & V_{z_k}(y_{N_y-1}) & B_y \\
T_y & 0 & \cdots & 0 & 0 & B_y^\dagger & V_{z_k}(y_{N_y})
\end{array}
\right]
\end{equation}
\end{widetext}
where $B_y$ is an $N_x$-dimensional matrix similar to $B_z$,
\begin{equation}
B_y=-\frac{1}{2h_y^2}I,
\end{equation}
and the block matrix $V_{z_k}(y_j)$ is defined on the $x$ line at the ($y_j,z_k$) point as
\begin{widetext}
\begin{equation}
\label{difference03}
V_{z_k}(y_j) = \left[
\begin{array}{ccccccc}
v_{y_j,z_k}(x_1) & b_x & 0 & 0 & \cdots & 0 & \tau_x^* \\
b_x^* & v_{y_j,z_k}(x_2) & b_x & 0 & & & 0 \\
0 & \ddots & \ddots & \ddots & \ddots & & \vdots \\
\vdots & \ddots & b_x^* & v_{y_j,z_k}(x_i) & b_x & \ddots & \vdots \\
\vdots & & \ddots & \ddots & \ddots & \ddots & 0 \\
0 & & & 0 & b_x^* & v_{y_j,z_k}(x_{N_x-1}) & b_x \\
\tau_x & 0 & \cdots & 0 & 0 & b_x^* & v_{y_j,z_k}(x_{N_x})
\end{array}
\right]
\end{equation}
\end{widetext}
with
\begin{equation}
b_x=-\frac{1}{2}c_{-1}= -\frac{1}{2}c_1=-\frac{1}{2h_x^2}
\end{equation}
and
\begin{equation}
v_{y_j,z_k}(x_i)= \frac{1}{h_x^2}+\frac{1}{h_y^2}+\frac{1}{h_z^2}+ v (x_i,y_j,z_k).
\end{equation}
Here, $T_y$ ($\tau_x$) is an $N_x$-dimensional matrix (a coefficient) determining the boundary condition in the $y$ ($x$) direction and $v$ is the sum of the external, Hartree, and exchange-correlation potentials. In the case when the periodic boundary conditions are imposed on all the directions,
\begin{equation}
T_z=\e^{ik_zL_z}B_z,
\label{eqn:zbound}
\end{equation}
\begin{equation}
T_y=\e^{ik_yL_y}B_y,
\end{equation}
and
\begin{equation}
\tau_x=\e^{ik_xL_x}b_x,
\end{equation}
where $k_x$, $k_y$, and $k_z$ are the Bloch wave numbers, and $L_x$, $L_y$, and $L_z$ are the unit-cell lengths. On the other hand, when the isolated boundary conditions are imposed, $T_z=0$, $T_y=0$, and $\tau_x=0$.

\newpage
\begin{center}
FIGURE CAPTIONS
\end{center}

\hspace{-0.5cm}FIG 1: Double-grid adopted in the text. The ``$\times$'' and ``$\bullet$'' correspond to coarse- and dense-grid points, respectively. The circle shows the core region of an atom which is taken to be sufficiently large to contain the cutoff region of nonlocal pseudopotentials.
\\

\hspace{-0.5cm}FIG 2: Functions on coarse- and dense-grid points in the one-dimensional case. $X_J$ ($x_j$) represents a coarse(dense)-grid point with $j=nJ+t$ $(0 \leq t < n)$, and so $X_J=x_{nJ}$. (a) The inner product between the wave function $\psi(x)$ and the nonlocal part of pseudopotential $v(x)$ evaluated over coarse-grid points. (b) The inner product evaluated over dense-grid points.
\\

\hspace{-0.5cm}FIG 3: Convergence of the total energy for the hydrogen atom as a function of the coarse-grid cutoff energy $E_c^{coars}$. The atom is located at the center between adjacent coarse-grid points.
\\

\hspace{-0.5cm}FIG 4: Variation of the total energy as a function of the displacement of the flourine atom relative to coarse-grid points along a coordinate axis. The coarse-grid spacing $h_\mu$ is 0.33 a.u.
\\

\hspace{-0.5cm}FIG 5: Relative positions with respect to the grids for F$_2$ and Cu$_2$ molecules and adiabatic potential curves. (a) The center of gravity of the molecule is placed at the center of neighboring grids with respect to the $x, y$ and $z$ directions. In the figure, $``\times"$ indicates the coarse grid and $\bullet$ represents nuclei. (b) The center of gravity of the molecule is placed on the grid plane which orthogonally intersects the coupling axis. For the grid plane parallel to the coupling axis, the center of gravity is placed at the center of neighboring grids. (c) Adiabatic potential curves of F$_2$ molecules. The grid spacing is 0.33 a.u. (d) Adiabatic potential curve for Cu$_2$ molecules. The grid spacing is 0.27 a.u. The calculated points are fit to spline functions as a guide to the eye.
\\

\hspace{-0.5cm}FIG 6: Relative errors between the Hartree potential obtained using boundary values computed by each multipole expansion method and that evaluated by the analytical solution for the cross section at $z=$0.075 a.u. Results using boundary values (a) obtained by the FCD-MPE method and (b) obtained by the multipole expansion around one point. Spheres show the locations of the nuclei.
\\

\hspace{-0.5cm}FIG 7: (a) Front and (b) axial views of the $(5,5)$ carbon nanotube. $r$, $\varphi$, $L_z$ are the radius, the helical twist angle and the translational shift of the nanotube, respectively.
\\

\hspace{-0.5cm}FIG 8: (a) Electronic band structure of the (5, 5) carbon nanotube using the supercell with length $L_z$, (b) pseudoband structure, and (c) electronic band structure using the conventional supercell with the length 2$L_z$. The zero of energy is chosen to be the Fermi level.


\begin{references}
\bibitem{comment}
The $q_{\mu\nu}$ is so defined as to be a coefficient of the second differentiation $\frac{\partial^2}{\partial \mu \partial \nu}\frac{1}{|\vecvar{r}|}$.
\end{references}
\end{document}